\documentclass[preprint,12pt]{elsarticle}

\usepackage{amssymb}
\usepackage{amsmath}
\usepackage{url}

\begin{document}

\begin{frontmatter}



\title{Economic complexity and regional development in India: Insights from a state-industry bipartite network}

\author[label1]{Joel M Thomas}

\author[label1,label2]{ Abhijit Chakraborty\corref{cor1}} 
\cortext[cor1]{abhijit@labs.iisertirupati.ac.in}
\affiliation[label1]{organization={Department of Humanities and Social Sciences,
Indian Institutes of Science Education and Research Tirupati},
            addressline={Srinivasapuram, Yerpedu Mandal}, 
            city={Tirupati},
            postcode={517619}, 
            state={Andhra Pradesh},
            country={India}}
\affiliation[label2]{organization={RIKEN Center for Interdisciplinary Theoretical and Mathematical Sciences},
            addressline={2-1 Hirosawa},
            city={Wako},
            postcode={351-0198},
            state={Saitama},
            country={Japan}}



\begin{abstract}

This study investigates the economic complexity of Indian states by constructing a state-industry bipartite network using firm-level data on registered companies and their paid-up capital. We compute the Economic Complexity Index and apply the fitness--complexity algorithm to quantify the diversity and sophistication of productive capabilities across the Indian states and two union territories. The results reveal substantial heterogeneity in regional capability structures, with states such as Maharashtra, Karnataka, and Delhi exhibiting consistently high complexity, while others remain concentrated in ubiquitous, low-value industries. The analysis also shows a strong positive relationship between complexity metrics and per-capita Gross State Domestic Product, underscoring the role of capability accumulation in shaping economic performance. Additionally, the number of active firms in India demonstrates a persistent exponential growth at an annual rate of $11.2\%$, reflecting ongoing formalization and industrial expansion. The ordered binary matrix displays the characteristic triangular structure observed in complexity studies, validating the applicability of complexity frameworks at the sub-national level. This work highlights the usefulness of firm-based data for assessing regional productive structures and emphasizes the importance of capability-oriented strategies for fostering balanced and sustainable development across Indian states. By demonstrating the usefulness of firm registry data in data constrained environments, this study advances the empirical application of economic complexity methods and provides a quantitative foundation for capability-oriented industrial and regional policy in India.

\end{abstract}




\end{frontmatter}



\section{Introduction}
\label{intro}
 Since Adam Smith’s seminal pin factory illustration, wealth has been linked to the specialization, and division of labour and knowledge. Societal skill grows through specialization, as individuals hold different tacit knowledge, which leads to diversification at higher levels such as firms, cities, and countries. Contrary to common belief, economies with more specialized individuals are more diversified, a fact supported by both classical theory~\cite{smith2002inquiry} and empirical evidence~\cite{bettencourt2014professional, petralia2017climbing}.
Quite naturally economic activities are unevenly distributed across geographical regions, reflecting differences in productive structures, institutional capacity, and historical development pathways. Understanding these variations and interactions of geographical regions is crucial for explaining long-term growth patterns and for designing effective regional development policies. The economic complexity method, introduced by Hidalgo and Hausmann~\cite{hidalgo2009building} and based on the relatedness~\cite{hidalgo2007product} between specific activities and locations, has proven to be highly effective in characterizing the detailed structure of economies~\cite{hidalgo2021economic}.

Unlike traditional growth theories, relatedness and complexity methods do not isolate individual factors but instead infer their combined presence without strong assumptions. It uses dimensionality reduction technique to uncover the combinations of factors that best explain the spatial distribution of economic activities. A central insight of the economic complexity is that the geographic pattern of production itself contains rich information. This information is captured by representing which products are made in which countries as a country-product matrix, where each entry indicates whether a country produces a given product above a specified threshold~\cite{hidalgo2009building, hausmann2014atlas, tacchella2012new, mariani2015measuring, caldarelli2012network, mealy2019interpreting}.
While this framework has been widely applied at the national level, recent studies have demonstrated its usefulness at sub-national scales, including analyses of U.S. metropolitan areas~\cite{mealy2019interpreting, fritz2021economic},  Brazil~\cite{operti2018dynamics}, Mexico~\cite{chavez2017economic}, China~\cite{gao2018quantifying}, Japanese prefectures~\cite{chakraborty2020economic}, and Australian states~\cite{reynolds2018sub}. These works highlight that regions-like countries-accumulate capabilities in distinct ways, shaping their ability to participate in complex and high-value industries. Measures of economic complexity have been shown to have strong predictive power across a wide range of outcomes, including future economic growth~\cite{hidalgo2009building, stojkoski2016impact, tacchella2018dynamical}, income inequality~\cite{hartmann2017linking, zhu2020export}, technological progress~\cite{nast2024fueling} and environmental impacts such as greenhouse gas emissions~\cite{neagu2019relationship, romero2021economic}.

India presents a particularly compelling case for sub-national complexity analysis due to its substantial interstate disparities in industrial diversification, income levels, and economic outcomes. Despite rapid national growth over the past three decades, development has been uneven, with a few states emerging as diversified industrial hubs while others remain dependent on a narrow range of low-complexity activities. Existing measures of state-level development often fail to fully capture the underlying structure of productive capabilities that drive these differences. In this context, economic complexity offers a more structural and capability-based perspective, enabling a deeper examination of regional divergence.

This study applies economic complexity methods to the Indian context by constructing a state--industry network using firm-level data from the Ministry of Corporate Affairs (MCA). Unlike studies based on trade flows or employment, this approach leverages the distribution of registered companies and their paid-up capital to infer capability patterns across states. We compute the Economic Complexity Index (ECI)~\cite{hidalgo2009building} and the fitness-complexity metrics~\cite{tacchella2012new} to quantify the diversity and sophistication of each state’s productive structure, and we analyse how these measures relate to per-capita Gross State Domestic Product (GSDP). By doing so, this work provides new insights into the structural foundations of regional development in India and demonstrates the utility of firm-based complexity metrics in data-constrained environments.Our study offers a detailed characterization of the economic structures of Indian states and establishes a quantitative foundation to inform industrial policy.

\section{Data}
\label{data}

The primary dataset used in this study is the \textit{Registrar of Companies (RoC) wise company master data} obtained from the MCA through the open government data platform of India~\cite{datagov}. The dataset contains detailed firm-level information for all firms registered under various RoCs across the country, including their corporate identification number, industrial classification according to the National Industrial Classification (NIC) system~\cite{nic}, date of incorporation, authorized capital, and paid-up capital. For the purposes of this analysis, only active firms as of the extraction date (10 July 2025) are retained to ensure that the network reflects the current productive landscape. Industries are classified at the two-digit NIC level, and sectors corresponding to the ``00'' category are excluded as they denote unclassified or miscellaneous activities. After filtering, the dataset covers $99$ distinct industrial categories across $27$ Indian states and the union territories of Delhi and Jammu \& Kashmir. We provide the state codes along with the number of active firms in each state in SI text 1. 

Paid-up capital is used as the measure of a state's investment in each industry, as it captures the actual equity contributed by shareholders and provides a consistent firm-level indicator in the absence of uniform data on revenues or employment. For each pair of state-industry, we aggregate the paid-up capital of all active firms to construct the weighted matrix $x_{sp}$, which forms the basis for computing the Revealed Comparative Advantage (RCA)~\cite{balassa1965trade}. The RCA quantifies whether a state is relatively more specialized in a particular industry by comparing the industry's share in the state's total paid-up capital with its corresponding share at the national level:
\[
RCA_{sp} =
\frac{x_{sp} / \sum_{p} x_{sp}}
{\left( \sum_{s} x_{sp} \right) / \left( \sum_{s,p} x_{sp} \right)} .
\]
The binary matrix $M_{sp}$, which underpins all complexity calculations, is derived by assigning $M_{sp} = 1$ when $\mathrm{RCA}_{sp} \ge 1$, and $M_{sp} = 0$ otherwise.

To relate the complexity measures, data on per-capita GSDP for the fiscal year 2023--24 are obtained from the Ministry of Statistics and Programme Implementation (MoSPI). These dataset enable an assessment of the relationship between productive capabilities and economic performance across states. Together, the MCA firm registry and MoSPI economic indicators provide a comprehensive foundation for examining structural differences in regional productive capacities within India.

\section{Research Methodology}
\label{method}
We employ both the economic complexity method and the fitness–complexity method. Below, we provide a brief description of each approach.

\subsection{Economic Complexity}

We construct a bipartite state--industry network using firm-level data as described in Section~\ref{data}. The network is represented by a binary adjacency matrix \( M_{sp} \), where \( M_{sp}=1 \) if state \( s \) is significantly present in industry \( p \), and \( M_{sp}=0 \) otherwise. Based on this matrix, we compute the Economic Complexity Index (ECI) following the Reflections Method introduced in Ref.~\cite{hidalgo2009building}.

The first step consists of computing the diversification of states and the ubiquity of industries, defined respectively as
\begin{equation}
k_{s,0} = \sum_{p} M_{sp}, \qquad
k_{p,0} = \sum_{s} M_{sp}.
\end{equation}
Here, \( k_{s,0} \) measures the number of industries in which state \( s \) is active, while \( k_{p,0} \) measures the number of states in which industry \( p \) is present. In network science, $k_{s,1}$ and $k_{p,1}$ correspond to the average nearest-neighbor degree.
Higher-order measures of economic complexity are obtained through an iterative procedure, referred to as reflections, which alternates between states and industries. The \(N\)-th order reflections are defined recursively as
\begin{align}
k_{s,N} &= \frac{1}{k_{s,0}} \sum_{p} M_{sp} \, k_{p,N-1}, \\
k_{p,N} &= \frac{1}{k_{p,0}} \sum_{s} M_{sp} \, k_{s,N-1}.
\end{align}
These relations capture the idea that a state is economically complex if it is connected to industries that are themselves rare, while an industry is complex if it is present only in diversified states.

By substituting the expression for \( k_{p,N-1} \) into the equation for \( k_{s,N} \), the reflections method can be rewritten as a linear transformation acting solely on states:
\begin{equation}
k_{s,N} = \sum_{s'} \widetilde{M}_{ss'} \, k_{s',N-2},
\end{equation}
where the transformed state--state matrix \( \widetilde{M} \) is defined as
\begin{equation}
\widetilde{M}_{ss'} =
\sum_{p}
\frac{M_{sp} \, M_{s'p}}{k_{s,0} \, k_{p,0}}.
\end{equation}

Similarly, an industry--industry transformed matrix can be constructed as
\begin{equation}
\widetilde{M}_{pp'} =
\sum_{s}
\frac{M_{sp} \, M_{sp'}}{k_{p,0} \, k_{s,0}}.
\end{equation}

The matrix \( \widetilde{M}_{ss'} \) quantifies the similarity between states in terms of their industrial structures, normalized by state diversification and industry ubiquity. The spectral properties of  \( \widetilde{M}_{ss'} \)  encode the structure of the bipartite network~\cite{mealy2019interpreting}.

The Economic Complexity Index (ECI) for states is obtained from the eigenvector associated with the second largest eigenvalue of the matrix \( \widetilde{M}_{ss'} \). The eigenvector corresponding to the largest eigenvalue represents a trivial uniform solution and is therefore discarded. The resulting eigenvector \( K_s \) is standardized to obtain
\begin{equation}
ECI_s = \frac{K_s - \langle K \rangle}{\mathrm{stdev}(K)}.
\end{equation}

An analogous procedure applied to the matrix \( \widetilde{M}_{pp'} \) yields the Product (or Industry) Complexity Index.

\subsection{Fitness complexity}
To complement the ECI results, we compute state fitness and industry complexity using the nonlinear fitness--complexity algorithm~\cite{tacchella2012new}. The method rests on three principles: (i) a prefecture’s fitness is determined by the diversity of its industrial sectors, weighted by their complexity; (ii) sectors present in many prefectures are considered less complex; and (iii) the complexity of a sector is ultimately constrained by the lowest-fitness prefectures that possess it. Mathematically, this is expressed through a set of self-consistent, coupled iterative equations defining the fitness $F_s$ of states and the complexity  $Q_p$ of industrial sectors.
\[
F_s^{(n)} = \sum_{p} M_{sp} \, Q_p^{(n-1)},
\]
\[
Q_p^{(n)} = \left( \sum_{s} \frac{M_{sp}}{F_s^{(n-1)}} \right)^{-1}.
\]
At each step, the values are normalized:
\[
F_s^{(n)} \leftarrow \frac{F_s^{(n)}}{\langle F^{(n)} \rangle}, \qquad
Q_p^{(n)} \leftarrow \frac{Q_p^{(n)}}{\langle Q^{(n)} \rangle}.
\]
Here, $n$ denotes an arbitrary iteration step.

Finally, we correlate the computed complexity indicators (ECI and fitness) with per-capita GSDP to evaluate how productive capabilities relate to economic performance across Indian states. We also report the ECI and fitness values of the states in SI Text 3, and the PCI values for the five highest- and lowest-ranked sectors in SI Text 4.

\section{Results}
\label{results}
\begin{figure}[t]
\centering
\includegraphics[width=\linewidth]{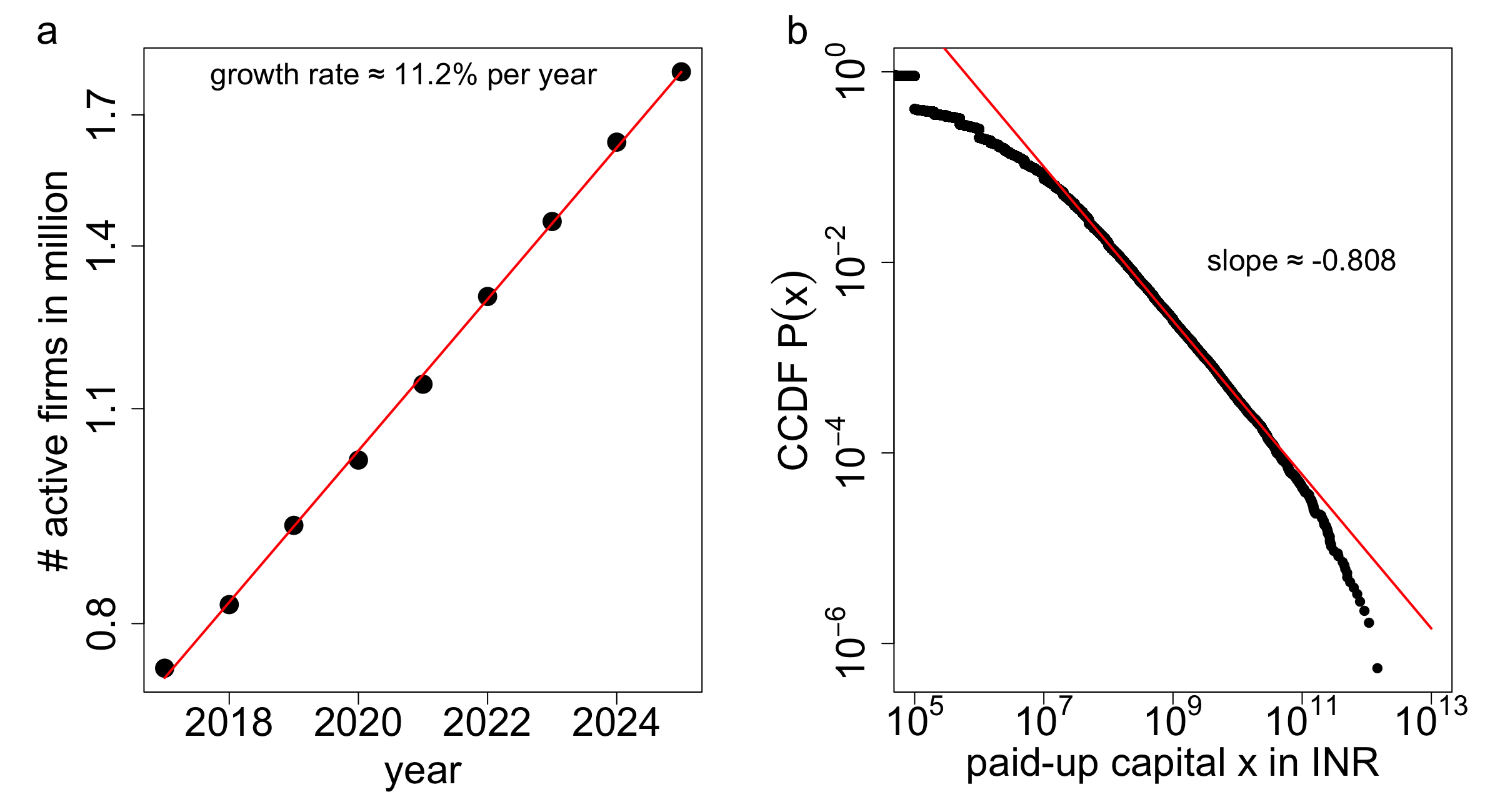}
\caption{ {\bf Growth dynamics and firm size distribution in India.}
(a) Number of active firms $N (t)$ in each year $t$. Points represent empirical observations, the red line denotes an exponential fit of the form $N(t) \sim e^{0.112t}$, indicating sustained growth at an annual rate of approximately $11.2\%$.
(b) Complementary cumulative distribution function (CCDF) $P(x)$ of the paid-up capital $x$, illustrating the firm-size distribution. The red line represents a power-law fit $P(x) \sim x^{-(\alpha+1)}$ in the intermediate range, with an estimated exponent $\alpha = -1.808$.
}
\label{fig1}
\end{figure}

We illustrate the temporal evolution of the number of active firms $N$ in India in Fig.~\ref{fig1}~(a), which reveals a striking exponential growth pattern $N(t) \sim e^{0.112t}$ over the past several years $t$. The number of active firms increases with a sustained growth at an annual rate of approximately $11.2\%$, and the semi-logarithmic plot exhibits a strong linear trend, confirmed by the exceptionally high correlation between year and log-transformed firm counts (Pearson correlation coefficient $ r = 0.99 , p = 7 \times 10^{-12}$). This exponential expansion is consistent with broader trends documented in emerging economies that undergo rapid structural transformation, where reductions in entry barriers, expansion of formal credit, and digitalization of business registration systems contribute to accelerated firm formation. In the Indian context, reforms such as the MCA21 electronic registry, the growth of the start-up ecosystem, and increasing formalization of economic activity have likely contributed to this sustained increase in registered firms. The pronounced growth of active firms thus reflects not only the increase in entrepreneurial activity, but also the deepening of industrial and financial structures, forming an essential empirical backdrop for subsequent analyzes of economic complexity at the state level.

The Complementary cumulative distribution function (CCDF) $P(x)$ of the paid-up capital $x$ in Fig.~\ref{fig1}~(b), revealing a highly skewed pattern characteristic of firm-size distributions observed in many real-world economies~\cite{pareto1964cours, okuyama1999zipf, axtell2001zipf}. The distribution exhibits a clear power-law behavior $P(x) \sim x^{-(\alpha+1)}$ in the intermediate range, with an estimated exponent of $\alpha = -1.806$, indicating that a relatively small number of firms account for a disproportionately large share of total capitalization. Such heavy-tailed distributions are widely documented in the literature~\cite{pareto1964cours, okuyama1999zipf, axtell2001zipf} on firm growth and income distribution, where mechanisms such as preferential attachment, heterogeneous capabilities, and scale-dependent investment thresholds generate broad size dispersion. The presence of a power law in our data suggests that India’s firm population is similarly shaped by strong underlying heterogeneity, with most firms operating at modest capital levels while a minority of highly capitalized firms dominate the financial landscape. This structural pattern reinforces the importance of using measures like paid-up capital in assessing economic complexity, as it highlights how variation in firm capitalization contributes to differences in industrial depth across states.
\begin{figure}[t]
\centering
\includegraphics[width=0.8\linewidth]{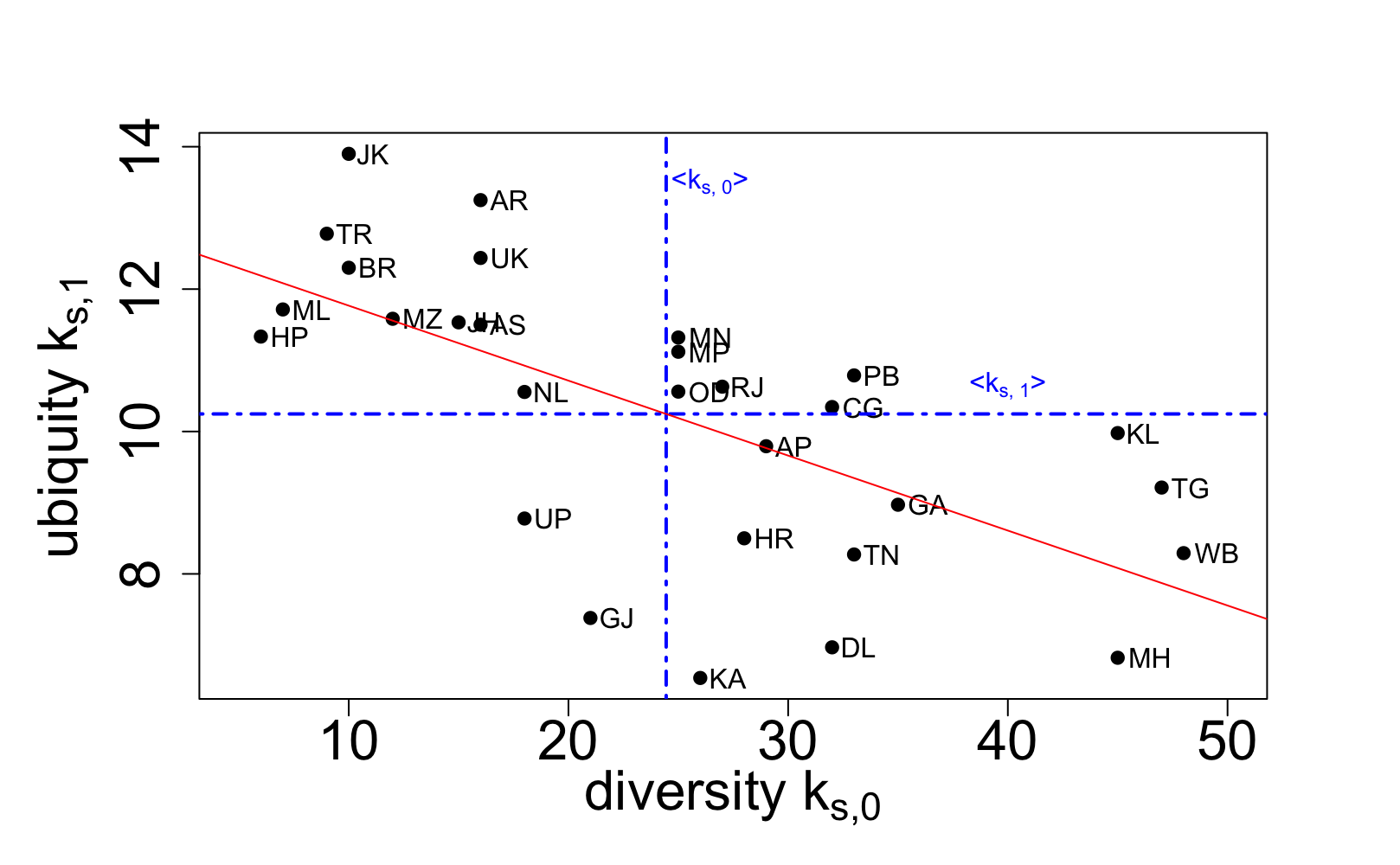}
\caption{ {\bf Diversification–ubiquity relationship across Indian states.}
States are positioned across four quadrants in $k_{s,0}-k_{s,1}$ plane based on their industrial diversity $k_{s,0}$ and the ubiquity $k_{s,1}$ of the industries in which they are active. The red line indicates a linear regression fit. The Pearson correlation coefficient is $r = 0.652$ with a corresponding $p$-value of $6 \times 10^{-4}$. The codes used for the states are provided in the SI Text 1. 
}
\label{fig2}
\end{figure}

The relationship between the diversification of Indian states (measured by $k_{s,0}$, the number of industries in which a state exhibits revealed comparative advantage) and ubiquity of its industrial sector (measured by $k_{s,1}$, the average number of states in which these industries are present) is shown in Fig.~\ref{fig2}. Consistent with previous studies~\cite{hidalgo2009building, gao2018quantifying, chakraborty2020economic} on economic complexity, the figure reveals a clear negative correlation between diversification and ubiquity, indicating that states engaged in a wider range of industrial activities tend to specialize in industries that are relatively uncommon across the country. This pattern mirrors the findings of Hidalgo and Hausmann~\cite{hidalgo2009building} for international trade networks, where more complex economies export a broad set of products that few other countries can produce. The linear fit in Fig.~\ref{fig2} underscores this inverse relationship, suggesting that Indian states follow similar structural regularities: highly diversified states such as Maharashtra, Karnataka, and Delhi are positioned in the quadrant associated with low industry ubiquity, while less diversified states such as Himachal Pradesh, Meghalaya, Tripura, Bihar and Jharkhand rely on more common, widely distributed industries. This empirical structure forms the basis for estimating the Economic Complexity Index (ECI) and highlights the uneven distribution of productive capabilities across Indian states. Further, we quantify the similarity between states and visualize their interrelationships using a minimum spanning tree, as presented in Supplementary Information Text 2.  

\begin{figure}[t]
\centering
\includegraphics[width=0.8\linewidth]{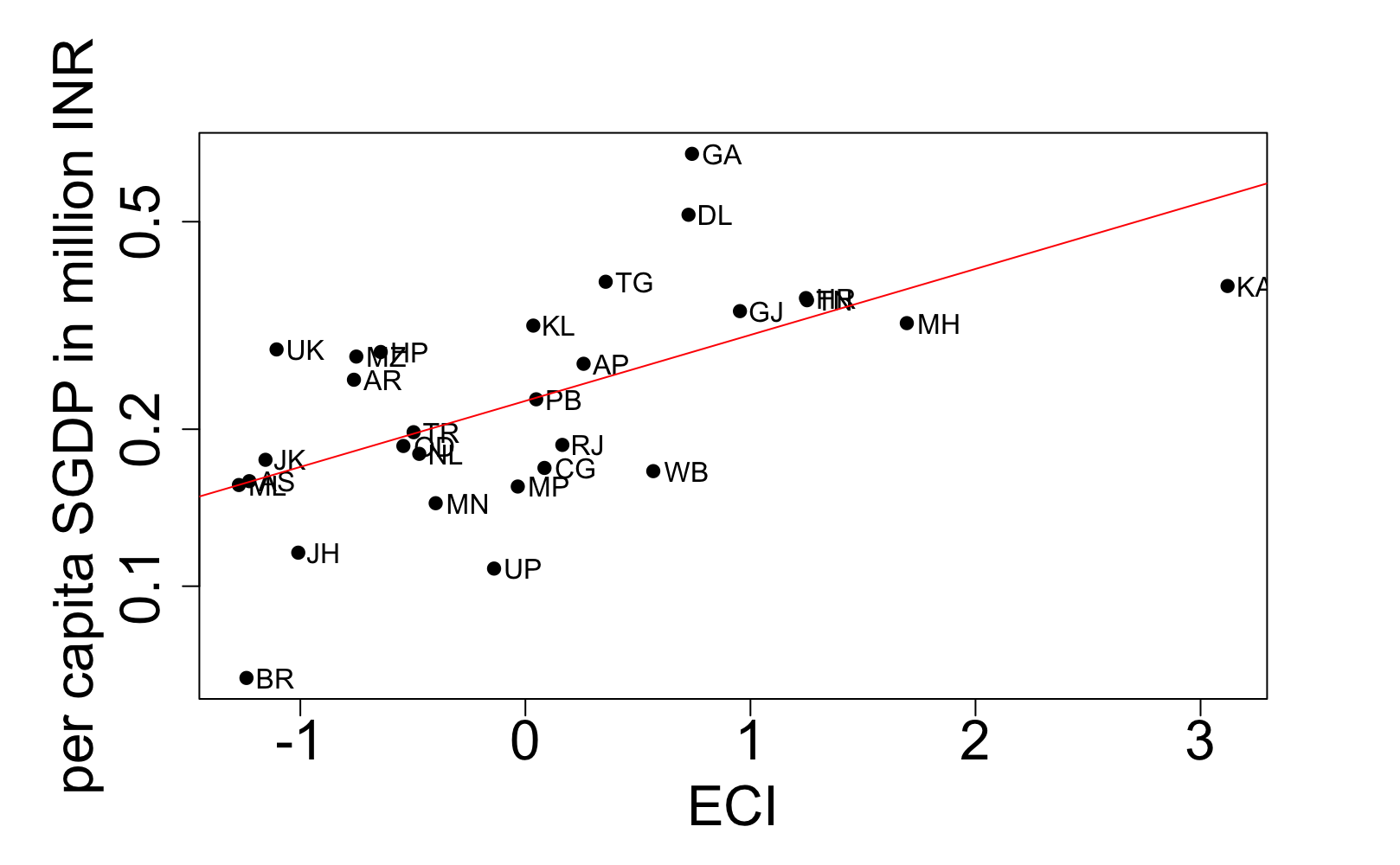}
\caption{{\bf Relationship between economic complexity and income across Indian states.} Per capita state gross domestic product vs ECI for the year 2023-24. The straight line represent an exponential fit to the graph.
States lying below the fitted trend exhibit lower income levels relative to their complexity, suggesting latent growth potential. The Pearson correlation coefficients between the two variables is found $r=0.599$ with $p-value = 6\times10^{-4}$.}
\label{fig3}
\end{figure}

We illustrates the relationship between the ECI of Indian states and their per-capita Gross State Domestic Product (GSDP) in Fig.~\ref{fig3} for the fiscal year $2023 - 24$. The scatter plot reveals a clear positive association between economic complexity and income levels, with the fitted exponential curve capturing this upward trend. States with higher ECI values, reflecting more diverse and less ubiquitous industrial capabilities tend to exhibit substantially higher per-capita GSDP, consistent with the core prediction of economic complexity theory that productive capability accumulation drives long-run economic prosperity. The strength of this relationship is supported by the Pearson correlation between ECI and log-transformed per-capita GSDP ($r = 0.599$, $p = 6 \times 10^{-4}$), indicating a statistically significant linkage between structural complexity and development outcomes. This aligns with global findings for countries and sub-national regions, reinforcing the view that states endowed with diversified industrial ecosystems and specialized capabilities achieve higher levels of economic performance. Here, we also observe that states such as West Bengal, Chhattisgarh, Madhya Pradesh, and Uttar Pradesh lie below the expected values of per-capita GSDP. This suggests that these states may have the potential for more rapid growth in the future.

\begin{figure}[t]
\centering
\includegraphics[width=0.98\linewidth]{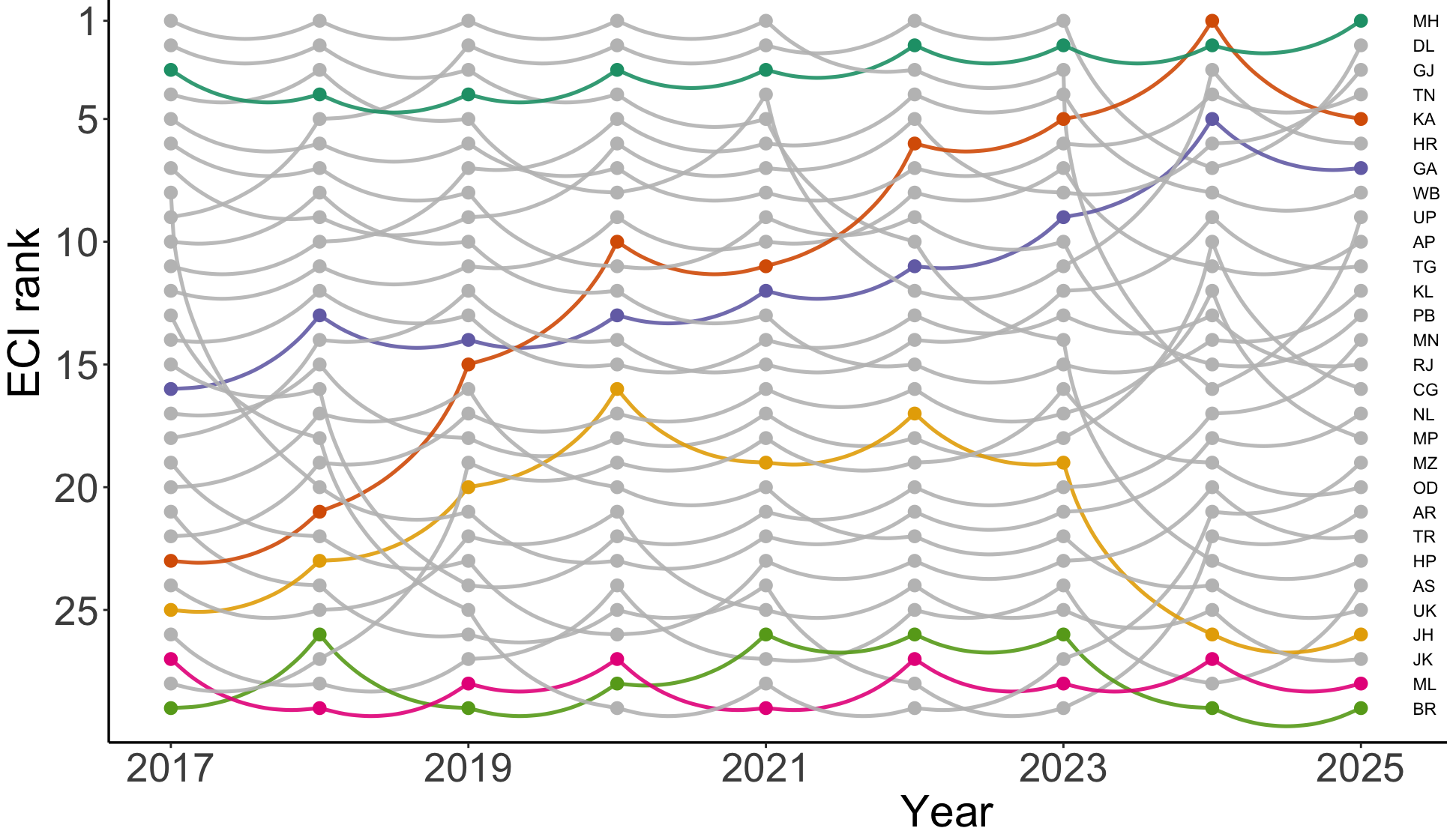}
\caption{{\bf Temporal evolution of state economic complexity rankings.} Plot showing the evolution of Economic Complexity Index (ECI) rankings of Indian states over time. States are ranked from highest (top) to lowest (bottom) complexity in each year. Thin gray lines represent all states, while selected states are highlighted to illustrate distinct developmental trajectories. Persistent high rankings of Maharashtra and Delhi indicate long-standing diversified capability bases, whereas the upward mobility of Karnataka reflects successful accumulation of complex capabilities. In contrast, the declining or stagnant ranks of several states underscore uneven regional capability development. Overall, the figure reveals both stability and mobility in India’s sub-national complexity hierarchy.}
\label{fig4}
\end{figure}
The evolution of the ECI rankings of Indian states over time is shown in Fig.~\ref{fig4}, which highlighting both long-term stability and notable shifts in regional productive capabilities. It shows that states such as Maharashtra and Delhi consistently occupy the highest positions in the ranking, reflecting their long-established industrial diversity and strong capability base. In contrast, states like Bihar and Meghalaya remain persistently at the lower end of the ranking, indicating limited diversification and dependence on more ubiquitous, low-complexity industries. The temporal trajectories also reveal important transitions: Karnataka and Goa exhibit significant upward movement in recent years, suggesting successful expansion into high-complexity sectors, while Uttar Pradesh, despite early high rankings, declines over time as other states develop more specialized capabilities. These dynamic shifts underscore the uneven pace at which productive knowledge accumulates across Indian states and emphasize that regional economic complexity is shaped by both historical industrial structures and ongoing capability development. As shown in Fig.~\ref{fig1}(a), the number of firms used to compute the ECI of the states is limited for the earlier years; however, it still provides a clear indication of the trend in the evolution of ECI rankings. Note that we use the financial year rather than the calendar year. For instance, the year 2017 includes all active firms registered before April 1, 2017, for the calculation of the ECI. 

\begin{figure}[t]
\centering
\includegraphics[width=0.99\linewidth]{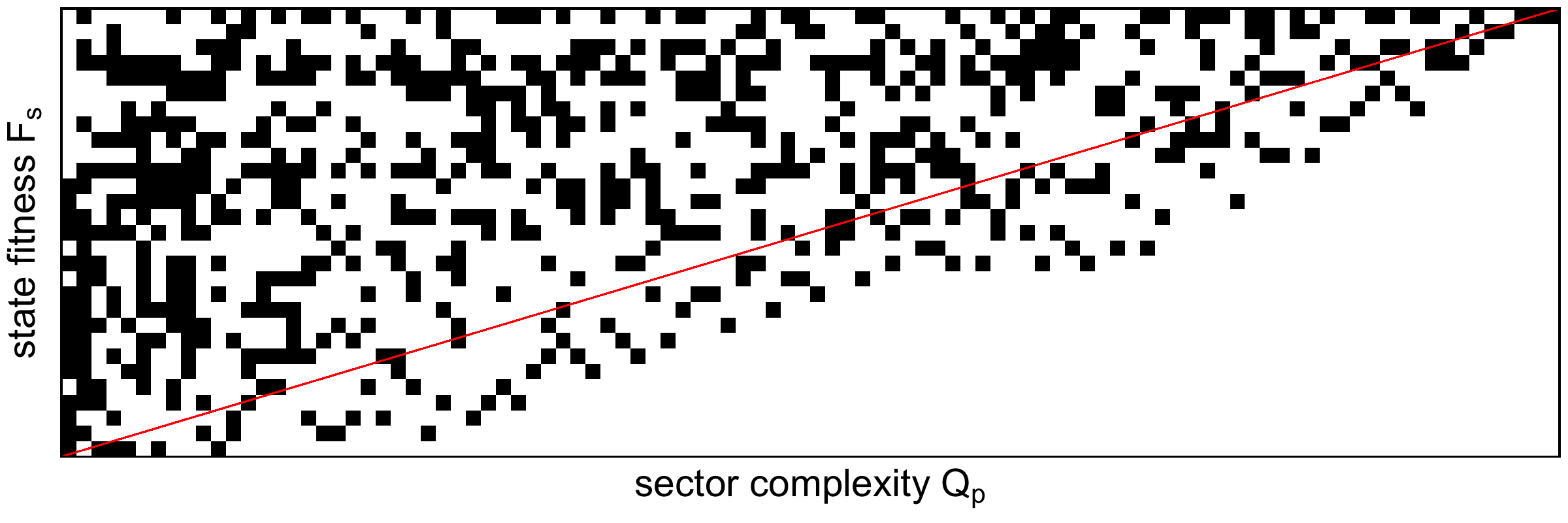}
\caption{{\bf Ordered state-industry matrix revealing hierarchical capability structure.} Binary state–industry matrix 
$M_{sp}$, ordered by decreasing state fitness (top to bottom) and increasing industry complexity (left to right), as obtained from the nonlinear fitness-complexity algorithm. Black cells indicate the presence of a revealed comparative advantage. The emergence of a pronounced triangular structure indicates that high-fitness states are active across both ubiquitous and complex industries, whereas low-fitness states participate only in highly ubiquitous sectors. 
}
\label{fig5}
\end{figure}

To assess the robustness of the ECI, we compare it with the results derived from the fitness-complexity method~\cite{tacchella2012new}. A brief description of this methodology is provided in Section~\ref{method}.
Fig.~\ref{fig5} displays the ordered binary state-industry matrix $M_{sp}$, arranged according to the rankings produced by the fitness-complexity algorithm. After ordering states by decreasing fitness and industries by increasing complexity, the matrix reveals a characteristic triangular structure, a well-known empirical signature in fitness-complexity analyzes. In this structure, highly fit states (those at the top of the matrix) tend to be active across a wide range of industries including both simple and highly complex ones while low fitness states (toward the bottom) participate only in the most ubiquitous and least complex industries. This pattern indicates that the accumulation of productive capabilities is hierarchical: advanced states possess the broad set of capabilities required to operate in sophisticated industries, whereas less-developed states are confined to capabilities associated with more common, low-complexity sectors. The emergence of this triangular shape thus validates the internal consistency of the fitness-complexity framework for Indian states and highlights the heterogeneous nature of industrial capabilities across the country.

\begin{figure}[t]
\centering
\includegraphics[width=0.9\linewidth]{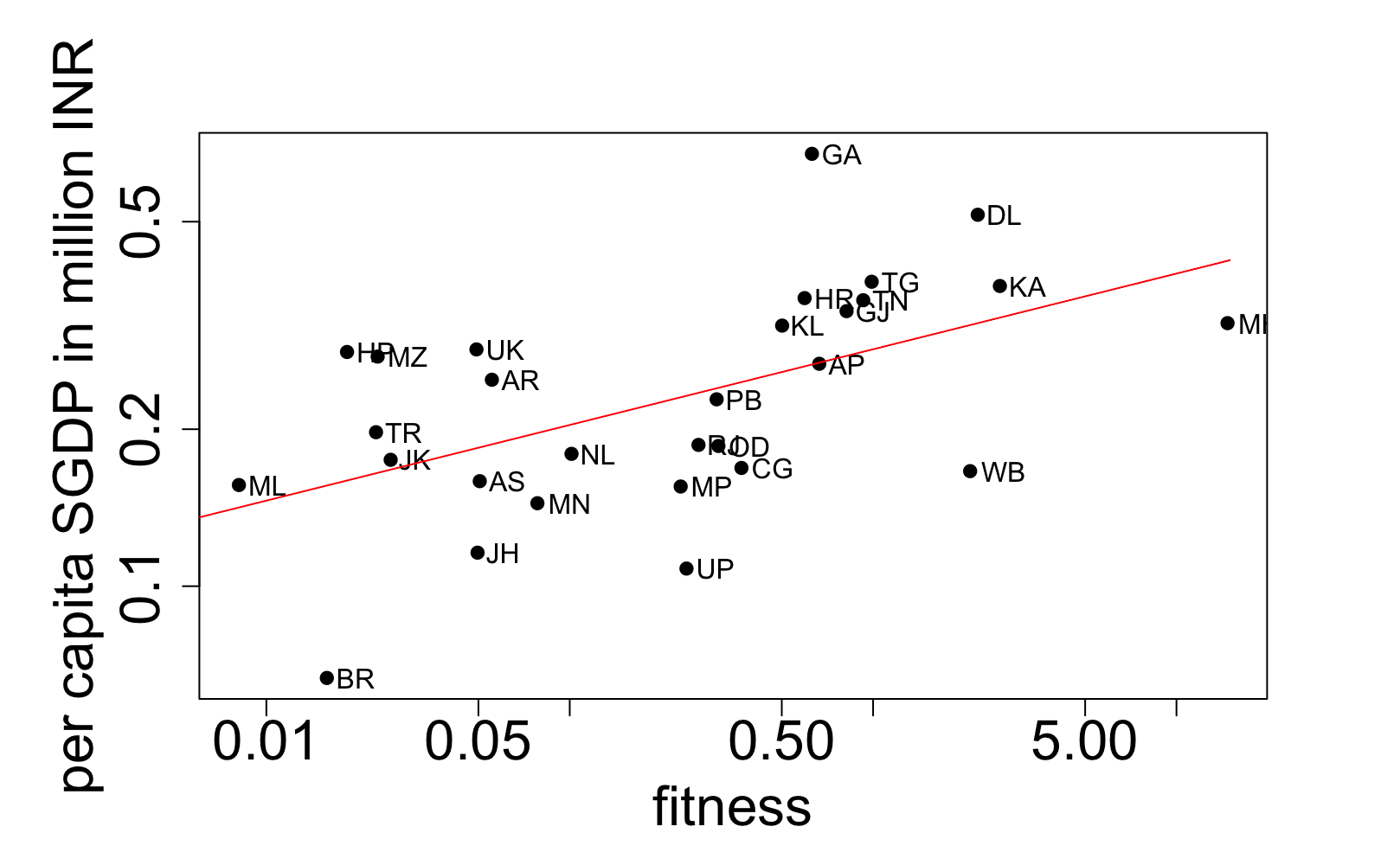}
\caption{ {\bf Relationship between state fitness and income levels.} Per capita state gross domestic product vs fitness for the year 2023-24. The straight line in the graph represent the power law fit that indicates states with broader and more sophisticated productive capability sets tend to achieve higher income levels.. The Pearson correlation coefficients between the two $\log$ variables is found  $r=0.5362$ and $p-value= 2.7\times10^{-3}$.}
\label{fig6}
\end{figure}

We examine the relationship between the fitness of Indian states derived from the nonlinear fitness-complexity algorithm and their per-capita GSDP in Fig.~\ref{fig6} for the year $2023-24$. The scatter plot reveals a clear positive association between these two variables, indicating that states with higher fitness scores tend to achieve higher income levels. The fitted power-law curve captures this trend, consistent with the theoretical expectation that fitness, which emphasizes the breadth and sophistication of productive capabilities, is strongly linked to long-term economic performance. The relationship is statistically significant, with the correlation between $\log(\text{fitness})$ and $\log(\text{per-capita GSDP})$ estimated at $r = 0.5362$ ($p = 2.7 \times 10^{-3}$), confirming that even the nonlinear fitness metric provides meaningful explanatory power for regional economic disparities. Similar to Fig.~\ref{fig3} it also indicates that states such as West Bengal, Chhattisgarh, Madhya Pradesh, and Uttar Pradesh fall below the expected per-capita GSDP levels, suggesting that they may possess greater potential for faster economic growth in the future.
\section{Discussions and conclusions}

This study investigates the economic complexity of Indian states by constructing a state-industry bipartite network using firm-level data on registered companies and their paid-up capital. Our analysis contributes to the growing body of sub-national economic complexity literature by demonstrating that firm registration data-despite its limitations-captures meaningful structural differences in productive capabilities across regions. Several noteworthy insights emerge from the findings.

First, the exponential increase in the number of active firms over time reflects India’s broader processes of economic formalization, administrative modernization, and expansion of industrial activity. Digital initiatives such as the MCA21 registry, reductions in entry barriers, and the rise of the start-up ecosystem have collectively enabled an unprecedented scale of firm creation. This trend strengthens the empirical foundation for applying complexity-based metrics at the state level, as a richer firm population enhances the resolution of capability patterns within the state-industry network.

Second, our results show that the fundamental structural relationships predicted by economic complexity theory-namely, the negative association between state diversification and industry ubiquity-hold robustly in the Indian context. Highly diversified states tend to specialize in industries that few others possess, reflecting deeper and more advanced capability sets. This structural regularity validates the use of both the ECI and fitness-complexity frameworks at the sub-national scale.

Third, both complexity measures exhibit strong correlations with per-capita GSDP, demonstrating that states with richer, more sophisticated productive ecosystems tend to achieve higher income levels. This finding aligns with international evidence from countries such as Japan, Australia, and the United States, and highlights the importance of capability accumulation in shaping regional development outcomes. The evolution of complexity rankings over time further reveals that while some states-such as Maharashtra and Delhi-maintain consistently high complexity, others, including Karnataka and Goa, have made notable strides, indicating dynamic capability growth. Conversely, persistently low-complexity states underline the structural challenges that constrain diversification and growth.

The ordered binary matrix generated by the fitness-complexity algorithm provides additional insight, revealing the characteristic triangular structure associated with capability hierarchies. This pattern underscores that productive capabilities develop cumulatively: high-fitness states possess the broad set of foundational capabilities required to enter complex sectors, while low-fitness states remain concentrated in ubiquitous, low-value industries.

Taken together, these findings suggest that economic complexity offers a powerful and informative framework for understanding regional economic disparities in India. The results point to several policy implications: states aspiring to enhance their economic performance should prioritize capability-building interventions, such as improving industrial infrastructure, fostering innovation ecosystems, strengthening vocational and technical skills, and enabling technology diffusion. Importantly, policies that merely promote sectoral expansion without addressing underlying capability constraints are unlikely to generate sustained complexity growth.

In conclusion, this work demonstrates that firm-level registry data, combined with complexity-based analytical tools, can yield valuable insights into the structural foundations of regional development. While future research could incorporate alternative measures of firm size-such as employment, revenues, or asset valuations-to complement paid-up capital, the present analysis significantly advances our understanding of productive capabilities within Indian states. As richer datasets become available, economic complexity can serve as a critical instrument for diagnosing regional strengths, identifying latent opportunities, and informing capability-driven development strategies across India.




\end{document}